\documentclass[10pt]{article}

\usepackage{a4wide}
\usepackage{amsmath,amssymb}
\usepackage{graphics}
\usepackage{pst-all}

\title{\textbf{On the Reeh-Schlieder Property in Curved Spacetime}}
\author{Ko Sanders\thanks{E-mail: jas503@york.ac.uk}\\
Department of Mathematics, University of York\\ 
Heslington, York YO10 5DD, United Kingdom.}
\date{January 29, 2008}

\newtheorem{definition}{Definition}[section]
\newtheorem{theorem}[definition]{Theorem}
\newtheorem{proposition}[definition]{Proposition}
\newtheorem{corollary}[definition]{Corollary}
\newtheorem{lemma}[definition]{Lemma}
\newenvironment{proof*}{\smallskip\par\noindent\emph{Proof. }
 \ignorespaces}{\hfill$\Box$\smallskip\par\ignorespaces}

\newcommand{\map}[3]{\ensuremath{#1\!:\!#2\!\rightarrow\!#3}}

\newcommand{\id}{\ensuremath{\mathrm{id}}}

\begin{document}
\maketitle
${}$\\[-1.6cm]
\begin{center}
{\it Dedicated to Klaas Landsman,\\
out of gratitude for the support he offered when it was most
needed.}\\[1.5cm]
\end{center}

\begin{abstract}
We attempt to prove the existence of Reeh-Schlieder states on curved
spacetimes in the framework of locally covariant quantum field theory
using the idea of spacetime deformation and assuming the existence of
a Reeh-Schlieder state on a diffeomorphic (but not isometric) spacetime.
We find that physically interesting states with a weak form of the
Reeh-Schlieder property always exist and indicate their usefulness.
Algebraic states satisfying the full Reeh-Schlieder property also
exist, but are not guaranteed to be of physical interest.
\end{abstract}

\section{Introduction}

The Reeh-Schlieder theorem (\cite{Reeh+}) is a result in axiomatic
quantum field theory which states that for a scalar Wightman field in
Minkowski spacetime
any state in the Hilbert space can be approximated arbitrarily well by
acting on the vacuum with operations performed in any prescribed open
region. The physical meaning of this is that the vacuum state has very
many non-local correlations and an experimenter in any given region can
exploit the vacuum fluctuations by performing a suitable measurement
in order to produce any desired state up to arbitrary accuracy.

In this paper we will investigate whether we can find states of a
quantum field system in a curved spacetime which have the same property,
(the Reeh-Schlieder property). We do this using the technique of
spacetime deformation, as pioneered in \cite{Fulling+} and as applied
successfully to prove a spin-statistics theorem in curved spacetime in
\cite{Verch1}. This means that we assume the existence of a
Reeh-Schlieder state (i.e. a state with the Reeh-Schlieder property) in
one spacetime and try to derive the existence of another state in a
diffeomorphic (but not isometric) spacetime which also has the
Reeh-Schlieder property. We will prove that for every given region there
is a state in the physical state space that has the Reeh-Schlieder
property for that particular region (but maybe not for all regions).
Algebraic states with the full Reeh-Schlieder property also exist, i.e.
states which have the Reeh-Schlieder property for all open regions
simultaneously. However, their existence follows from an abstract
existence principle and, consequently, such states are not guaranteed
to be of any physical interest.

To keep the discussion as general as possible we will work in the
axiomatic language known as locally covariant quantum field theory
as introduced in \cite{Brunetti+} (see also \cite{Verch1}, where some
of these ideas already appeared, and \cite{Brunetti+2} for a recent
application). We outline this formulation in section \ref{LCQFTsection}
and our most important assumption there will be the time-slice axiom,
which expresses the existence of a causal dynamical law. In section
\ref{deformationsection} we will prove the geometric results on
spacetime deformation that we need and we will see what they mean for a
locally covariant quantum field theory. Section \ref{RSsection} contains
our main results on deforming one Reeh-Schlieder state into another one
and it notes some immediate consequences regarding the type of local
algebras and Tomita-Takesaki modular theory. As an example we discuss
the free scalar field in section \ref{freefieldsection} and we end with
a few conclusions.

\section{Locally covariant quantum field theory}\label{LCQFTsection}

In this section we briefly describe the main ideas of locally covariant
quantum field theory as introduced in \cite{Brunetti+}. It will also
serve to fix our notation for the subsequent sections.

In the following any quantum physical system will be described by a
$C^*$-algebra $\mathcal{A}$ with a unit $I$, whose self-adjoint elements
are the observables of the system. It will be advantageous to consider a
whole class of possible systems rather than just one.
\begin{definition}\label{def_alg}
The category $\mathfrak{Alg}$ has as its objects all unital $C^*$-algebras
$\mathcal{A}$ and as its morphisms all injective $^*$-homomorphisms
$\alpha$ such that $\alpha(I)=I$. The product of morphisms is given by
the composition of maps and the identity map
$\id_{\mathcal{A}}$ on a given object serves as an identity morphism.
\end{definition}
A morphism $\map{\alpha}{\mathcal{A}_1}{\mathcal{A}_2}$ expresses the fact
that the system described by $\mathcal{A}_1$ is a sub-system of that
described by $\mathcal{A}_2$, which is called a super-system. The
injectivity of the morphisms means that, as a matter of principle, any
observable of a sub-system can always be measured, regardless of any
practical restrictions that a super-system may impose.

A state of a system is represented by a normalised positive linear
functional $\omega$, i.e. $\omega(A^*A)\ge 0$ for all $A\in\mathcal{A}$
and $\omega(I)=1$. The set of all states on $\mathcal{A}$ will be
denoted by $\mathcal{A}^{*+}_1$. Not all of these states are of physical
interest, so it will be convenient to have the following notion at our
disposal.
\begin{definition}\label{def_states}
The category $\mathfrak{States}$ has as its objects all subsets
$S\subset\mathcal{A}^{*+}_1$, for all unital $C^*$-algebras
$\mathcal{A}$ in $\mathfrak{Alg}$, which are closed under convex linear
combinations and under operations from $\mathcal{A}$ (i.e.
$\frac{\omega(A^*.A)}{\omega(A^*A)}\in S$ if $\omega\in S$ and
$A\in\mathcal{A}$ such that $\omega(A^*A)\not=0$) and as its morphisms
all maps
$\map{\alpha^*}{S_1}{S_2}$ for which $S_i\subset(\mathcal{A}_i)^{*+}_1$,
$i=1,2$, and $\alpha^*$ is the restriction of the dual of a morphism
$\map{\alpha}{\mathcal{A}_2}{\mathcal{A}_1}$ in $\mathfrak{Alg}$,
i.e. $\alpha^*(\omega)=\omega\circ\alpha$ for all $\omega\in S_1$. Again
the product of morphisms is given by the composition of maps and the
identity map $\id_S$ on a given object serves as an identity morphism.
\end{definition}


After these operational aspects we now turn to the physical ones. The
systems we will consider are intended to model quantum fields living in
a (region of) spacetime which is endowed with a fixed Lorentzian metric
(a background gravitational field). The relation between sub-systems
will come about naturally by considering sub-regions of spacetime. More
precisely we consider the following:
\begin{definition}\label{def_man}
By the term \emph{globally hyperbolic spacetime} we will mean a
connected, Hausdorff, paracompact, $C^{\infty}$ Lorentzian manifold
$M=(\mathcal{M},g)$ of dimension $d=4$, which is oriented, time-oriented
and globally hyperbolic.

A subset $O\subset\mathcal{M}$ of a globally hyperbolic spacetime $M$ is
called \emph{causally convex} iff for all $x,y\in O$ all causal curves
from $x$ to $y$ lie entirely in $O$. A non-empty open set which is
connected and causally convex is called a causally convex region or
\emph{cc-region}. A cc-region whose closure is compact is called a
\emph{bounded cc-region}.

The category $\mathfrak{Man}$ has as its objects all globally hyperbolic
spacetimes $M=(\mathcal{M},g)$ and its morphisms $\Psi$ are given by all
maps $\map{\psi}{\mathcal{M}_1}{\mathcal{M}_2}$ which are smooth
isometric embeddings (i.e.
$\map{\psi}{\mathcal{M}_1}{\psi(\mathcal{M}_1)}$ is a diffeomorphism and
$\psi_*g_1=g_2|_{\psi(\mathcal{M}_1)}$) such that the orientation and
time-orientation are preserved and $\psi(\mathcal{M}_1)$ is causally
convex. Again the product of morphisms is given by the composition of
maps and the identity map $\id_M$ on a given object serves as a unit.
\end{definition}
A region $O$ in a globally hyperbolic spacetime is causally convex if
and only if $O$ itself is globally hyperbolic (see \cite{Hawking+}
section 6.6), so a cc-region is exactly a connected globally hyperbolic
region.

The image of a morphism is by definition a cc-region. Notice that the
converse also holds. If $O\subset\mathcal{M}$ is a cc-region then
$(O,g|_O)$ defines a globally hyperbolic spacetime in its own right. In
this case there is a canonical morphism $I_{M,O}:O\rightarrow M$ given
by the canonical embedding $\map{\iota}{O}{\mathcal{M}}$. We will often
drop $I_{M,O}$ and $\iota$ from the notation and simply write
$O\subset M$.

The importance of causally convex sets is that for any morphism $\Psi$
the causality structure of $M_1$ coincides with that of $\Psi(M_1)$ in
$M_2$:
\begin{equation}\label{subscript}
\psi(J_{M_1}^{\pm}(x))=J_{M_2}^{\pm}(\psi(x))
\cap\psi(\mathcal{M}_1),\quad x\in\mathcal{M}_1.
\end{equation}
If this were not the case then the behaviour of a quantum physical
system living in $\mathcal{M}_1$ could depend in an essential way on the
super-system, which makes it practically impossible to study the smaller
system as a sub-system in its own right. This possibility is
therefore excluded from in mathematical framework.

Equation (\ref{subscript}) allows us to drop the subscript in
$J^{\pm}_M$ if we introduce the convention that $J^{\pm}$ is always
taken in the largest spacetime under consideration. This simplifies the
notation without causing any confusion, even when
$O\subset M_1\subset M_2$ with canonical embeddings, because then we
just have $J^{\pm}(O):=J^{\pm}_{M_2}(O)$ and
$J^{\pm}_{M_1}(O)=J^{\pm}(O)\cap\mathcal{M}_1$. Similarly we take by
convention
\begin{eqnarray}
D(O)&:=&D_{M_2}(O),\nonumber\\
O^{\perp}&:=&O^{\perp_{M_2}}:=\mathcal{M}_2\setminus\overline{J(O)},
\nonumber
\end{eqnarray}
and we deduce from causal convexity that
$D_{M_1}(O)=D(O)\cap\mathcal{M}_1$ and
$O^{\perp_{M_1}}=O^{\perp}\cap\mathcal{M}_1$.

The following lemma gives some ways of obtaining causally convex
sets in a globally hyperbolic spacetime.
\begin{lemma}\label{subspacetimes}
Let $M=(\mathcal{M},g)$ be a globally hyperbolic spacetime, 
$O\subset\mathcal{M}$ an open subset and $A\subset\mathcal{M}$
an achronal set.
Then:
\begin{enumerate}
\item the intersection of two causally convex sets is causally convex,
\item for any subset $S\subset M$ the sets $I^{\pm}(S)$ are causally
convex,
\item $O^{\perp}$ is causally convex,
\item $O$ is causally convex iff $O=J^+(O)\cap J^-(O)$,
\item $\mathrm{int}(D(A))$ and $\mathrm{int}(D^{\pm}(A))$ are causally
convex,
\item if $O$ is a cc-region, then $D(O)$ is a cc-region,
\item if $S\subset M$ is an acausal continuous hypersurface then
$D(S)$, $D(S)\cap I^+(S)$ and $D(S)\cap I^-(S)$ are open and causally
convex.
\end{enumerate}
\end{lemma}
\begin{proof*}
The first two items follow directly from the definitions. The fourth
follows from $J^+(O)\cap J^-(O)=\cup_{p,q\in O}(J^+(p)\cap J^-(q))$,
which is contained in $O$ if and only if $O$ is causally convex. The
fifth item follows from the first two and theorem 14.38 and lemma 14.6
in \cite{ONeill}.

To prove the third item, assume that $\gamma$ is a causal curve between
points in $O^{\perp}$ and $p\in\overline{J(O)}$ lies on $\gamma$. By
perturbing one of the endpoints of $\gamma$ in $O^{\perp}$ we may ensure
that the curve is time-like. Then we may perturb $p$ on $\gamma$ so that
$p\in\mathrm{int}(J(O))$ and $\gamma$ is still causal. This gives a
contradiction, because there then exists a causal curve from $O$ through
$p$ to either $x$ or $y$.

For the sixth statement we let $S\subset O$ be a smooth Cauchy surface
for $O$ (see \cite{Bernal+1}) and note that $D(O)$ is non-empty,
connected and $D(O)=D(S)$. The causal convexity of $O$ implies that
$S\subset\mathcal{M}$ is acausal, which reduces this case to statement
seven. The first part of statement seven is just lemma 14.43 and theorem
14.38 in \cite{ONeill}. The rest of statement seven follows from
statement one and two together with the openness of $I^{\pm}(S)$.
\end{proof*}

We now come to the main set of definitions, which combine the notions
introduced above (see \cite{Brunetti+}).
\begin{definition}\label{def_lcqft}
A \emph{locally covariant quantum field theory} is a covariant functor
$\map{\mathbf{A}}{\mathfrak{Man}}{\mathfrak{Alg}}$, written as
$M\mapsto \mathcal{A}_M$, $\Psi\mapsto\alpha_{\Psi}$.

A \emph{state space} for a locally covariant quantum field theory
$\mathbf{A}$ is a contravariant functor
$\map{\mathbf{S}}{\mathfrak{Man}}{\mathfrak{States}}$, such that for all
objects $M$ we have $M\mapsto S_M\subset(\mathcal{A}_M)^{*+}_1$ and for
all morphisms $\map{\Psi}{M_1}{M_2}$ we have
$\Psi\mapsto \alpha_{\Psi}^*|_{S_{M_2}}$. The set $S_M$ is called the
\emph{state space} for $M$.
\end{definition}

When it is clear that $\Psi=I_{M,O}$ for a canonical embedding
$\iota:O\rightarrow\mathcal{M}$ of a cc-region $O$ in a globally
hyperbolic spacetime $\mathcal{M}$, i.e. when $O\subset M$, we will
often simply write $\mathcal{A}_O\subset\mathcal{A}_M$ instead of using
$\alpha_{I_{M,O}}$. For a morphism $\map{\Psi}{M}{M'}$ which restricts
to a morphism $\map{\Psi|_O}{O}{O'\subset M}$ we then have
\begin{equation}\label{embedding}
\alpha_{\Psi|_O}=\alpha_{\Psi}|_{\mathcal{A}_O}
\end{equation}
rather than $\alpha_{I_{M',O'}}\circ\alpha_{\Psi|_O}=
\alpha_{\Psi}\circ\alpha_{I_{M,O}}$, as one can see from a commutative
diagram.

The framework of locally covariant quantum field theory is a
generalisation of algebraic quantum field theory (see
\cite{Brunetti+,Haag}). We now proceed to discuss several physically
desirable properties that such a locally covariant quantum field theory
and its state space may have (cf. \cite{Brunetti+}, but note that our
time-slice axiom is stronger).
\begin{definition}
A locally covariant quantum field theory $\mathbf{A}$ is called
\emph{causal} iff for any two morphisms $\map{\Psi_i}{M_i}{M}$, $i=1,2$
such that $\psi_1(\mathcal{M}_1)\subset(\psi_2(\mathcal{M}_2))^{\perp}$
in $\mathcal{M}$ we have
$\left[\alpha_{\Psi_1}(\mathcal{A}_{M_1}),\alpha_{\Psi_2}
(\mathcal{A}_{M_2})\right]=\left\{0\right\}$ in $\mathcal{A}_M$.

A locally covariant quantum field theory $\mathbf{A}$ with state
space $\mathbf{S}$ satisfies the \emph{time-slice axiom} iff for all
morphisms $\map{\Psi}{M_1}{M_2}$ such that $\psi(\mathcal{M}_1)$
contains a Cauchy surface for $\mathcal{M}_2$ we have
$\alpha_{\Psi}(\mathcal{A}_{M_1})=\mathcal{A}_{M_2}$ and
$\alpha_{\Psi}^*(S_{M_2})=S_{M_1}$.

A state space $\mathbf{S}$ for a locally covariant quantum field theory
$\mathbf{A}$ is called \emph{locally quasi-equivalent} iff for every
morphism $\map{\Psi}{M_1}{M_2}$ such that
$\psi(\mathcal{M}_1)\subset\mathcal{M}_2$ is bounded and for
every pair of states $\omega,\omega'\in S_{M_2}$ the GNS-representations
$\pi_{\omega},\pi_{\omega'}$ of $\mathcal{A}_{M_2}$ are quasi-equivalent
on $\alpha_{\Psi}(\mathcal{A}_{M_1})$. The local von Neumann algebras
$\mathcal{R}^{\omega}_{M_1}:=\pi_{\omega}(\alpha_{\Psi}
(\mathcal{A}_{M_1}))''$ are then *-isomorphic for all
$\omega\in S_{M_2}$.

A locally covariant quantum field theory $\mathbf{A}$ with a state space
functor $\mathbf{S}$ is called \emph{nowhere classical} iff for every
morphism $\map{\Psi}{M_1}{M_2}$ and for every state $\omega\in S_{M_2}$
the local von Neumann algebra $\mathcal{R}^{\omega}_{M_1}$ is not
commutative.
\end{definition}
Note that the condition
$\psi_1(\mathcal{M}_1)\subset(\psi_2(\mathcal{M}_2))^{\perp}$ is
symmetric in $i=1,2$. The causality condition formulates how the quantum
physical system interplays with the classical gravitational background
field, whereas the time-slice axiom expresses the existence of a causal
dynamical law. The condition of a locally quasi-equivalent state space
is more technical in nature and means that all states of a system can be
described in the same Hilbert space representation as long as we only
consider operations in a small (i.e. bounded) cc-region of the spacetime.

The condition that $\psi(\mathcal{M}_1)$ contains a Cauchy surface for
$\mathcal{M}_2$ is equivalent to $D(\psi(\mathcal{M}_1))=\mathcal{M}_2$,
because a Cauchy surface $S\subset\mathcal{M}_1$ maps to a Cauchy
surface $\psi(S)$ for $D(\psi(\mathcal{M}_1))$. On the algebraic level
this yields:
\begin{lemma}\label{Cauchydevelopment}
For a locally covariant quantum field theory $\mathbf{A}$ with a state
space $\mathbf{S}$ satisfying the time-slice axiom, an object
$(\mathcal{M},g)\in\mathfrak{Man}$ and a cc-region $O\subset\mathcal{M}$
we have $\mathcal{A}_O=\mathcal{A}_{D(O)}$ and $S_O=S_{D(O)}$. If $O$
contains a Cauchy surface of $\mathcal{M}$ we have
$\mathcal{A}_O=\mathcal{A}_{\mathcal{M}}$ and $S_O=S_M$.
\end{lemma}
\begin{proof*}
Note that both $(O,g|_O)$ and $(D(O),g|_{D(O)})$ are objects of
$\mathfrak{Man}$ (by lemma \ref{subspacetimes}) and that a Cauchy
surface $S$ for $O$ is also a Cauchy surface for $D(O)$. (The causal
convexity of $O$ in $\mathcal{M}$ prevents multiple intersections of
$S$.) The first statement then reduces to the second. Leaving the
canonical embedding implicit in the notation, the result immediately
follows from the time-slice axiom.
\end{proof*}

Finally we define the Reeh-Schlieder property, which we will study in
more detail in the subsequent sections.
\begin{definition}\label{RSproperty}
Consider a locally covariant quantum field theory $\mathbf{A}$ with
a state space $\mathbf{S}$. A state $\omega\in S_M$ has the
\emph{Reeh-Schlieder property} for a cc-region $O\subset\mathcal{M}$ iff
\[
\overline{\pi_{\omega}(\mathcal{A}_O)\Omega_{\omega}}=
\mathcal{H}_{\omega}
\]
where $(\pi_{\omega},\Omega_{\omega},\mathcal{H}_{\omega})$ is the
GNS-representation of $\mathcal{A}_M$ in the state $\omega$. We then say
that $\omega$ is a \emph{Reeh-Schlieder state for $O$}. We say that
$\omega$ is a \emph{(full) Reeh-Schlieder state} iff it is a
Reeh-Schlieder state for all cc-regions in $\mathcal{M}$.
\end{definition}

\section{Spacetime deformation}\label{deformationsection}

The existence of Hadamard states of the free scalar field in certain
curved spacetimes was proved in \cite{Fulling+} by deforming Minkowski
spacetime into another globally hyperbolic spacetime. Using a similar
but slightly more technical spacetime deformation argument \cite{Verch1}
proved a spin-statistics theorem for locally covariant quantum field
theories with a spin structure, given that such a theorem holds in
Minkowski spacetime. In the next section we will assume the existence
of a Reeh-Schlieder state in one spacetime and try to deduce along
similar lines the existence of such states on a deformed spacetime.
As a geometric prerequisite we will state and prove in the present
section a spacetime deformation result employing similar methods as the
references mentioned above.

First we recall the spacetime deformation result due to \cite{Fulling+}:
\begin{proposition}\label{deformation1}
Consider two globally hyperbolic spacetimes $M_i$, $i=1,2$, with
spacelike Cauchy surfaces $C_i$ both diffeomorphic to $C$. Then there
exists a globally hyperbolic spacetime $M'=(\mathbb{R}\times C,g')$
with spacelike Cauchy surfaces $C'_i$, $i=1,2$, such that $C'_i$ is
isometrically diffeomorphic to $C_i$ and an open neighbourhood of $C'_i$
is isometrically diffeomorphic to an open neighbourhood of $C_i$.
\end{proposition}
The proof is omitted, because the stronger result proposition
\ref{deformation2} will be proved later on. Note, however, the following
interesting corollary (cf. \cite{Brunetti+} section 4):
\begin{corollary}\label{isomorphism1}
Two globally hyperbolic spacetimes $M_i$ with diffeomorphic Cauchy
surfaces are mapped to isomorphic $C^*$-algebras $\mathcal{A}_{M_i}$ by
any locally covariant quantum field theory $\mathbf{A}$ satisfying the
time-slice axiom (with some state space $\mathbf{S}$).
\end{corollary}
\begin{proof*}
Consider two diffeomorphic globally hyperbolic spacetimes $M_i$,
$i=1,2$, let $M'$ be the deforming spacetime of proposition
\ref{deformation1} and let $W_i\subset\mathcal{M}_i$ be open
neighbourhoods of the Cauchy surfaces $C_i\subset\mathcal{M}_i$ which
are isometrically diffeomorphic under $\psi_i$ to the open
neighbourhoods $W'_i\subset\mathcal{M}'$ of the Cauchy surfaces
$C'_i\subset\mathcal{M}'$. We may take the $W_i$ and $W'_i$ to be
cc-regions (as will be shown in proposition \ref{deformation2}), so
that the $\Psi_i$ (determined by $\psi_i$) are isomorphisms in
$\mathfrak{Man}$. It then follows from lemma \ref{Cauchydevelopment}
that
\begin{eqnarray}
\mathcal{A}_{M_1}&=&\mathcal{A}_{W_1}=\mathcal{A}_{\psi_1^{-1}(W_1')}=
\alpha_{\Psi_1}^{-1}(\mathcal{A}_{W_1'})=\alpha_{\Psi_1}^{-1}
(\mathcal{A}_{M'})\nonumber\\
&=&\alpha_{\Psi_1}^{-1}\circ\alpha_{\Psi_2}(\mathcal{A}_{M_2}),\nonumber
\end{eqnarray}
where the $\alpha_{\Psi_i}$ are $^*$-isomorphisms. This proves the
assertion.
\end{proof*}

At this point a warning seems in place. Whenever $g_1,g_2$ are two
Lorentzian metrics on a manifold $\mathcal{M}$ such that both
$M_i:=(\mathcal{M},g_i)$ are objects in $\mathfrak{Man}$, corollary
\ref{isomorphism1} gives a $^*$-isomorphism $\alpha$ between the algebras
$\mathcal{A}_{M_i}$. If $O\subset\mathcal{M}$ is a cc-region for $g_1$
then $\alpha$ is a $^*$-isomorphism from $\mathcal{A}_{(O,g_1)}$ into
$\mathcal{A}_{M_2}$. However, the image cannot always be identified with
$\mathcal{A}_{(O,g_2)}$, because $O$ need not be causally convex for
$g_2$, in which case the object is not defined.

\begin{figure}
\begin{center}
\psset{xunit=5mm,yunit=5mm,runit=5mm}
\begin{pspicture}(-13,-5)(13,4)
\psline(-12,-4)(-12,4)
\psline(-6,-4)(-6,4)
\psline(-3,-4)(-3,4)
\psline(3,-4)(3,4)
\psline(6,-4)(6,4)
\psline(12,-4)(12,4)

\psline(-12,1)(-6,1)
\psline(-3,1)(3,1)
\psline(-3,-1)(3,-1)
\psline(6,-1)(12,-1)

\psline(-11,1)(-10,2)
\psline(-10,2)(-8,2)
\psline(-8,2)(-7,1)
\psline(-2,1)(-1,2)
\psline(-1,2)(1,2)
\psline(1,2)(2,1)

\psline(-1,1)(0,2)
\psline(0,2)(1,1)
\psline(-10,1)(-9,2)
\psline(-9,2)(-8,1)

\psline(-2.5,-1)(-1.5,-2)
\psline(-1.5,-2)(1.5,-2)
\psline(1.5,-2)(2.5,-1)
\psline(6.5,-1)(7.5,-2)
\psline(7.5,-2)(10.5,-2)
\psline(10.5,-2)(11.5,-1)

\psline(-0.5,-1)(0,-1.5)
\psline(0,-1.5)(0.5,-1)
\psline(8.5,-1)(9,-1.5)
\psline(9,-1.5)(9.5,-1)

\psellipse(-9,0)(1.5,2)
\psline[linestyle=dotted](-2.5,-1)(-2.5,1)
\psline[linestyle=dotted](2.5,-1)(2.5,1)
\psbezier[linestyle=dashed](-2,1)(-2.5,0.5)(-1.9,-0.5)(-2.4,-1)
\psbezier[linestyle=dashed](2,1)(2.5,0.5)(1.9,-0.5)(2.4,-1)
\psbezier[linestyle=dashed](-0.5,-1)(-1,-0.5)(-0.4,0.5)(-0.9,1)
\psbezier[linestyle=dashed](0.5,-1)(1,-0.5)(0.4,0.5)(0.9,1)

\rput(-9,-4.5){$M_2$}
\rput(0,-4.5){$M'$}
\rput(9,-4.5){$M_1$}
\rput(-13,1){$C_2$}
\rput(-4,1){$C'_2$}
\rput(-11,3.5){$W_2$}
\rput(-2,3.5){$W'_2$}
\rput(-2,-3.5){$W'_1$}
\rput(7,-3.5){$W_1$}
\rput(-9,-1){$O_2$}
\rput(-11,2){$V_2$}
\rput(-9,2.5){$U_2$}
\rput(-9,0.5){$K_2$}
\rput(-11,0.5){$N_2$}
\rput(-4,-1){$C'_1$}
\rput(5,-1){$C_1$}
\rput(9,-0.5){$K_1$}
\rput(7,-0.5){$N_1$}
\rput(6.7,-2){$V_1$}
\rput(10,-1.5){$U_1$}
\end{pspicture}
\end{center}
\caption{Sketch of the geometry of proposition
\ref{deformation2}.}\label{fig1}
\end{figure}
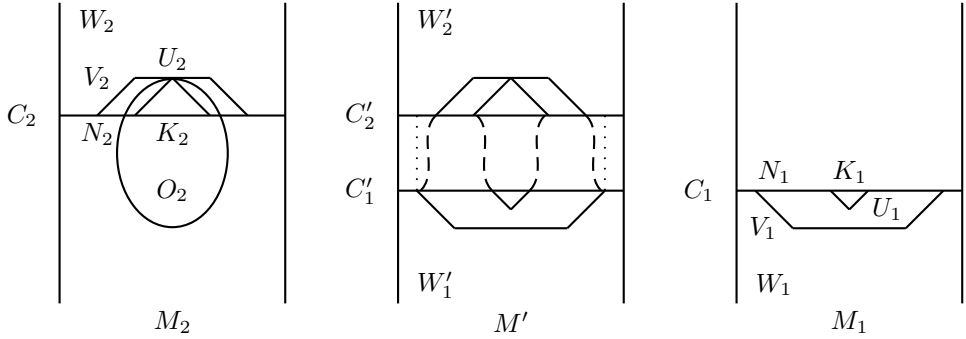

We now formulate and prove our deformation result. The geometric
situation is schematically depicted in figure \ref{fig1}.
\begin{proposition}\label{deformation2}
Consider two globally hyperbolic spacetimes $M_i$, $i=1,2$, with
diffeomorphic Cauchy surfaces and a bounded cc-region
$O_2\subset\mathcal{M}_2$ with non-empty causal complement,
$O_2^{\perp}\not=\emptyset$. Then there are a globally hyperbolic
spacetime $M'=(\mathcal{M}',g')$, spacelike Cauchy surfaces
$C_i\subset\mathcal{M}_i$ and $C_1',C_2'\in\mathcal{M}'$ and
bounded cc-regions $U_2,V_2\subset\mathcal{M}_2$ and
$U_1,V_1\subset\mathcal{M}_1$ such that the following hold:
\begin{itemize}
\item There are isometric diffeomorphisms $\map{\psi_i}{W_i}{W'_i}$
where $W_1:=I^-(C_1)$, $W'_1:=I^-(C'_1)$, $W_2:=I^+(C_2)$ and
$W'_2:=I^+(C'_2)$,
\item $U_2,V_2\subset W_2$, $U_2\subset D(O_2)$, $O_2\subset D(V_2)$,
\item $U_1,V_1\subset W_1$, $U_1\not=\emptyset$,
$V_1^{\perp}\not=\emptyset$, $\psi_1(U_1)\subset D(\psi_2(U_2))$
and $\psi_2(V_2)\subset D(\psi_1(V_1))$.
\end{itemize}
\end{proposition}
\begin{proof*}
First we recall the result of \cite{Bernal+1} that for any globally
hyperbolic spacetime $(\mathcal{M},g)$ there is a diffeomorphism
$\map{F}{\mathcal{M}}{\mathbb{R}\times C}$ for some smooth three
dimensional manifold $C$ in such a way that for each $t\in\mathbb{R}$
the surface $F^{-1}(\left\{t\right\}\times C)$ is a spacelike Cauchy
surface. The pushed-forward metric $g':=F_*g$ makes
$(\mathbb{R}\times C,g')$ a globally hyperbolic manifold, where $g'$
is given by
\begin{equation}\label{metric}
g'_{\mu\nu}=\beta dt_{\mu}dt_{\nu}-h_{\mu\nu}.
\end{equation}
Here $dt$ is the differential of the canonical projection on the
first coordinate $\map{t}{\mathbb{R}\times C}{\mathbb{R}}$, which is
a smooth time function; $\beta$ is a strictly positive smooth function
and $h_{\mu\nu}$ is a (space and time dependent) Riemannian metric on
$C$. The orientation and time orientation of $\mathcal{M}$ induce an
orientation and time orientation on $\mathbb{R}\times C$ via $F$. (If
necessary we may compose $F$ with the time-reversal diffeomorphism
$(t,x)\mapsto(-t,x)$ of $\mathbb{R}\times C$ to ensure that the function
$t$ increases in the positive time direction.) Applying the above to
the $M_i$ gives us two diffeomorphisms
$\map{F_i}{\mathcal{M}_i}{\mathcal{M}'}$, where
$\mathcal{M}'=\mathbb{R}\times C$ as a manifold. Note that we can take
the same $C$ for both $i=1,2$ by the assumption of diffeomorphic Cauchy
surfaces.

Define $O'_2:=F_2(O_2)$ and let $t_{\min}$ and $t_{\max}$ be the minimum
and maximum value that the function $t$ attains on the compact set
$\overline{O'_2}$. We now prove that
$F_2^{-1}((t_{\min},t_{\max})\times C)\cap O_2^{\perp}\not=\emptyset$.
Indeed, if this were empty, then we see that $\overline{J(O_2)}$
contains $F_2^{-1}(\left[t_{\min},t_{\max}\right]\times C)$ and hence
also $C_{\max}:=F_2^{-1}(\left\{t_{\max}\right\}\times C)$ and
$C_{\min}:=F_2^{-1}(\left\{t_{\min}\right\}\times C)$. In fact,
$C_{\min}\subset\overline{J^-(O_2)}$. Indeed, if
$p:=F_2^{-1}(t_{\min},x)$ is in $\overline{J^+(O_2)}$ then we can
consider a basis of neighbourhoods of $p$ of the form
$I^-(F_2^{-1}(t_{\min}+1/n,x))\cap
I^+(F_2^{-1}(\left\{t_{\min}-1/n\right\}\times C))$. If
$q_n\in J^+(O_2)$ is in such a basic neighbourhood, then the same
neighbourhood also contains a point $p_n\in O_2$. Hence, given a
sequence $q_n$ in $J^+(O_2)$ converging to $p$ we find a sequence $p_n$
in $O_2$ converging to $p$ and we conclude that
$p\in\overline{O_2}\subset\overline{J^-(O_2)}$. Similarly we can show
that $C_{\max}\subset\overline{J^+(O_2)}$. It then follows that
$I^+(C_{\max})\subset\overline{J^+(O_2)}$ and
$I^-(C_{\min})\subset\overline{J^-(O_2)}$, so that
$\overline{J(O_2)}=\mathcal{M}$ and $O^{\perp}=\emptyset$. This
contradicts our assumption on $O_2$, so we must have
$F_2^{-1}((t_{\min},t_{\max})\times C)\cap O_2^{\perp}\not=\emptyset$.
Then we may choose $t_2\in (t_{\min},t_{\max})$ such that
$C_2:=F_2^{-1}(\left\{t_2\right\}\times C)$ satisfies
$C_2\cap O_2\not=\emptyset$ and $C_2\cap O_2^{\perp}\not=\emptyset$. We
define $C'_2:=F_2(C_2)$, $W_2:=I^+(C_2)$ and
$W'_2:=(t_2,\infty)\times C$.

Note that $C_2\cap J(\overline{O_2})$ is compact. (We can
$\overline{O_2}$ by a finite number of open sets of the form
$I^{\pm}(q_n)$ and apply theorem 8.3.12 in \cite{Wald} to each point
$q_n$.) It follows that we can find relatively compact open sets
$K,N\subset C$ such that $K'_2:=\left\{t_2\right\}\times K$, $K_2:=F_2^{-1}(K'_2)$,
$N'_2:=\left\{t_2\right\}\times N$ and $N_2:=F_2^{-1}(N'_2)$ satisfy
$K\not=\emptyset$, $\overline{N}\not= C$, $\overline{K_2}\subset O_2$
and $C_2\cap J(\overline{O_2})\subset N_2$. We let
$C_{\max}:=F_2^{-1}(\left\{t_{\max}\right\}\times C)$ and define
$U_2:=D(K_2)\cap I^+(K_2)\cap I^-(C_{\max})$ and
$V_2:=D(N_2)\cap I^+(N_2)\cap I^-(C_{\max})$. It follows from lemma
\ref{subspacetimes} that $U_2,V_2$ are bounded cc-regions in $M_2$.
Clearly $U_2,V_2\subset W_2$, $U_2\subset D(O_2)$, $O_2\subset D(V_2)$
and $V_2^{\perp}\not=\emptyset$.

Next we choose $t_1\in (t_{\min},t_2)$ and define
$C'_1:=\left\{t_1\right\}\times C$, $C_1:=F_1^{-1}(C'_1)$,
$W_1:=I^-(C_1)$ and $W'_1:=(-\infty,t_1)\times C$. Let
$N',K'\subset C$ be relatively compact connected open sets such that
$K'\not=\emptyset$, $\overline{N'}\not=C$, $\overline{K'}\subset K$
and $\overline{N}\subset N'$. We define
$N'_1:=\left\{t_1\right\}\times N'$,
$K'_1:=\left\{t_1\right\}\times K'$, $N_1:=F_1^{-1}(N'_1)$,
$K_1:=F_1^{-1}(K'_1)$ and
$C_{\min}:=F_1^{-1}(\left\{t_{\min}\right\}\times C)$. Let
$U_1:=D(K_1)\cap I^-(K_1)\cap I^+(C_{\min})$ and
$V_1:=D(N_1)\cap I^-(N_1)\cap I^+(C_{\min})$. Again by lemma
\ref{subspacetimes} these are bounded cc-regions in $\mathcal{M}_1$.
Note that $U_1,V_1\subset W_1$ and $V_1^{\perp}\not=\emptyset$.

The metric $g'$ of $\mathcal{M}'$ is now chosen to be of the form
\[
g'_{\mu\nu}:=\beta dt_{\mu}dt_{\nu}-f\cdot (h_1)_{\mu\nu}-
(1-f)\cdot(h_2)_{\mu\nu}
\]
where we have written
$((F_i)_*g_i)_{\mu\nu}=\beta_i dt_{\mu}dt_{\nu}-(h_i)_{\mu\nu}$,
$f$ is a smooth function on $\mathcal{M}'$ which is identically $1$ on
$W'_1$, identically $0$ on $W'_2$ and $0<f<1$ on the intermediate region
$(t_1,t_2)\times C$ and $\beta$ is a positive smooth function which is
identically $\beta_i$ on $W'_i$. It is then clear that the maps $F_i$
restrict to isometric diffeomorphisms $\map{\psi_i}{W_i}{W'_i}$.

The function $\beta$ may be chosen small enough on the region 
$(t_1,t_2)\times C$ to make $(\mathcal{M},g')$ globally hyperbolic. (As
pointed out in \cite{Fulling+} in their proof of proposition
\ref{deformation1}, choosing $\beta$ small ``closes up'' the light cones and
prevents causal curves from ``running off to spatial infinity'' in the
intermediate region.) Furthermore, using the compactness of
$(t_1,t_2)\times N'$ and the continuity of $(h_i)_{\mu\nu}$ we see
that we may choose $\beta$ small enough on this set to ensure that any
causal curve through $\overline{K'_1}$ must also intersect $K'_2$ and
any causal curve through $\overline{N'_2}$ must also intersect $N'_1$.
This means that $\overline{K'_1}\subset D(K'_2)$ and
$\overline{N'_2}\subset D(N'_2)$ and hence
$\psi_1(U_1)\subset D(\psi_2(U_2))$ and
$\psi_2(V_2)\subset D(\psi_1(V_1))$. This completes the proof.
\end{proof*}

\begin{figure}
\begin{center}
\psset{xunit=5mm,yunit=5mm,runit=5mm}
\begin{pspicture}(-13,-5)(13,4)
\psline(-12,-4)(-12,4)
\psline(-2,-4)(-2,4)
\psline(2,-4)(2,4)
\psline(12,-4)(12,4)

\psline(-12,1)(-2,1)
\psline(-12,2)(-2,2)
\psline(2,1)(12,1)
\psline(2,2)(12,2)
\psline(2,-1)(12,-1)
\psline(2,-2)(12,-2)

\pscircle(-4.5,0.5){0.5}

\pspolygon*[linecolor=gray](-11,2)(-10.75,2.5)(-7.25,2.5)(-7,2)
\pspolygon[linecolor=black](-11,2)(-10.75,2.5)(-7.25,2.5)(-7,2)

\pspolygon*[linecolor=gray](3,2)(3.25,2.5)(6.75,2.5)(7,2)
\pspolygon[linecolor=black](3,2)(3.25,2.5)(6.75,2.5)(7,2)

\pspolygon*[linecolor=gray](-6,2)(-5.75,2.5)(-3.25,2.5)(-3,2)
\pspolygon[linecolor=black](-6,2)(-5.75,2.5)(-3.25,2.5)(-3,2)

\pspolygon*[linecolor=gray](8,2)(8.25,2.5)(10.75,2.5)(11,2)
\pspolygon[linecolor=black](8,2)(8.25,2.5)(10.75,2.5)(11,2)

\pspolygon*[linecolor=gray](4,-2)(4.25,-2.5)(5.75,-2.5)(6,-2)
\pspolygon[linecolor=black](4,-2)(4.25,-2.5)(5.75,-2.5)(6,-2)

\pspolygon*[linecolor=gray](7,-2)(7.25,-2.5)(11.75,-2.5)(12,-2)
\pspolygon[linecolor=black](7,-2)(7.25,-2.5)(11.75,-2.5)(12,-2)

\psline[linestyle=dashed](-3,2)(-4.5,-1)
\psline[linestyle=dashed](-6,2)(-4.5,-1)

\psline[linestyle=dashed](8,2)(7.5,1)
\psline[linestyle=dashed](7.5,1)(7.5,-1)
\psline[linestyle=dashed](7.5,-1)(7,-2)

\psline[linestyle=dashed](11,2)(11.5,1)
\psline[linestyle=dashed](11.5,1)(11.5,-1)
\psline[linestyle=dashed](12,-2)(11.5,-1)

\psline[linestyle=dashed](3,2)(3.5,1)
\psline[linestyle=dashed](3.5,1)(3.5,-1)
\psline[linestyle=dashed](4,-2)(3.5,-1)

\psline[linestyle=dashed](7,2)(6.5,1)
\psline[linestyle=dashed](6.5,1)(6.5,-1)
\psline[linestyle=dashed](6,-2)(6.5,-1)

\rput(-7,-4.5){$M_2$}
\rput(7,-4.5){$M'$}
\rput(-13,1){$C_2$}
\rput(-13,2){$T_2$}
\rput(1,1){$C'_2$}
\rput(1,2){$T'_2$}
\rput(1,-1){$C'_1$}
\rput(1,-2){$T'_1$}
\rput(-11,3.5){$W_2$}
\rput(3,3.5){$W'_2$}
\rput(3,-3.5){$W'_1$}
\rput(-4.5,0.5){$P_2$}
\rput(-4.5,3){$R$}
\rput(9.5,3){$R'$}
\rput(9.5,-3){$P'_1$}
\rput(5,-3){$Q'_1$}
\rput(5,3){$Q'_2$}
\rput(-9,3){$Q_2$}

\rput(-4.5,1.5){$N_2$}
\rput(9.5,-1.5){$N'_1$}
\rput(5,-1.5){$L'_1$}
\rput(-9,1.5){$L_2$}
\end{pspicture}
\end{center}
\caption{Sketch of the proof of the second part of 
proposition \ref{isomorphism2}.}\label{fig2}
\end{figure}
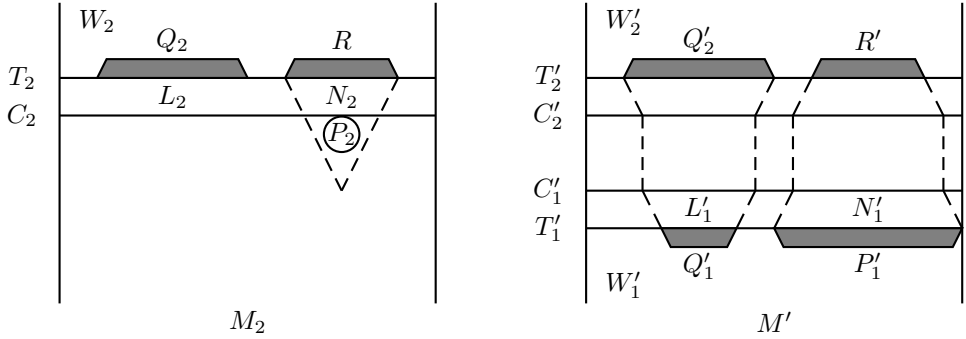

The analogue of corollary \ref{isomorphism1} for the situation of
proposition \ref{deformation2} is:
\begin{proposition}\label{isomorphism2}
Consider a locally covariant quantum field theory $\mathbf{A}$ with a
state space $\mathbf{S}$ satisfying the time-slice axiom and two
globally hyperbolic spacetimes $M_i$, $i=1,2$ with diffeomorphic Cauchy
surfaces. For any bounded cc-region $O_2\subset\mathcal{M}_2$ with
non-empty causal complement there are bounded cc-regions
$U_1,V_1\subset\mathcal{M}_1$ and a $^*$-isomorphism
$\map{\alpha}{\mathcal{A}_{M_2}}{\mathcal{A}_{M_1}}$ such that
$V_1^{\perp}\not=\emptyset$ and
\begin{equation}\label{estimate1}
\mathcal{A}_{U_1}\subset\alpha(\mathcal{A}_{O_2})\subset
\mathcal{A}_{V_1}.
\end{equation}

Moreover, if the spacelike Cauchy surfaces of the $M_i$ are non-compact
and $P_2\subset\mathcal{M}_2$ is any bounded cc-region, then there are
bounded cc-regions $Q_2\subset\mathcal{M}_2$ and
$P_1,Q_1\subset\mathcal{M}_1$ such that
$Q_i\subset P_i^{\perp}$ for $i=1,2$ and
\begin{equation}\label{estimate2}
\alpha(\mathcal{A}_{P_2})\subset\mathcal{A}_{P_1},\quad
\mathcal{A}_{Q_1}\subset\alpha(\mathcal{A}_{Q_2}),
\end{equation}
where $\alpha$ is the same $^*$-isomorphism as in the first part of this
proposition.
\end{proposition}
\begin{proof*}
We apply proposition \ref{deformation2} to obtain sets $U_i,V_i$ with
and isomorphisms $\Psi_i:W_i\rightarrow W'_i$ associated to the
isometric diffeomorphisms $\psi_i$. As in the proof of corollary
\ref{isomorphism1} the $\Psi_i$ give rise to $^*$-isomorphisms
$\alpha_{\Psi_i}$ and
$\alpha:=\alpha_{\Psi_1}^{-1}\circ\alpha_{\Psi_2}$ is a $^*$-isomorphism
from $\mathcal{A}_{M_2}$ to $\mathcal{A}_{M_1}$. Using the properties
of $U_i,V_i$ stated in proposition \ref{deformation2} we deduce:
\begin{eqnarray}
\mathcal{A}_{U_1}&=&\alpha_{\Psi_1}^{-1}(\mathcal{A}_{U'_1})\subset
\alpha_{\Psi_1}^{-1}(\mathcal{A}_{D(U'_2)})=
\alpha_{\Psi_1}^{-1}(\mathcal{A}_{U'_2})
=\alpha(\mathcal{A}_{U_2})\subset\alpha(\mathcal{A}_{O_2})\nonumber\\
&\subset&\alpha(\mathcal{A}_{V_2})=
\alpha_{\psi_1}^{-1}(\mathcal{A}_{V'_2})\subset
\alpha_{\psi_1}^{-1}(\mathcal{A}_{D(V'_1)})=
\alpha_{\psi_1}^{-1}(\mathcal{A}_{V'_1})=\mathcal{A}_{V_1}.\nonumber
\end{eqnarray}
Here we repeatedly used equation (\ref{embedding}) and lemma
\ref{Cauchydevelopment} (the time slice axiom). This proves the first
part of the proposition.

Now suppose that the Cauchy-surfaces are non-compact and let $P_2$ be any
bounded cc-region. We refer to figure \ref{fig2} for a depiction of this
part of the proof.

First choose Cauchy surfaces $T_2,T_+\subset W_2$ such that
$T_+\subset I^+(T_2)$. Note that
$J(\overline{P_2})\cap T_2$ is compact, so it has a relatively compact
connected open neighbourhood $N_2\subset T_2$. Choosing $T_+$
appropriately we see that $R:=D(N_2)\cap I^+(N_2)\cap I^-(T_+)$ is a
bounded cc-region in $\mathcal{M}_2$ by lemma \ref{subspacetimes} and
as usual we set $R':=\psi_2(R)$.

Now let $T'_-,T'_1\subset W_1'$ be Cauchy surfaces such that
$T'_-\subset I^-(T'_1)$ and note that $J(\overline{R'})\cap T'_1$ is
again compact, so we can find a relatively compact connected open
neighbourhood $N_1'\subset T'_1$ and use lemma \ref{subspacetimes} to
define the bounded cc-region
$P'_1:=D(N'_1)\cap I^-(N'_1)\cap I^+(T'_-)$ and
$P_1:=\psi_1^{-1}(P'_1)$.

Now let $L'_1\subset T'_1$ be a connected relatively compact set
such that $L'_1\cap N'_1=\emptyset$. Such an $L'_1$ exists because
$T'_1$ is non-compact. Define
$Q'_1:=D(L'_1)\cap I^-(L'_1)\cap I^+(T'_-)$ and
$Q_1:=\psi_1^{-1}(Q'_1)$. We see that $Q_1\subset P_1^{\perp}$ is
a bounded cc-region and $Q'_1\subset D(\psi_2(L_2))$ where
$L_2\subset T_2\setminus N$ is a relatively compact open set. In fact,
we can choose $L_2$ to be connected because $Q'_1$ lies in a connected
component $C$ of $D(\psi_2(T_2\setminus N))$. We now define the
bounded cc-region $Q_2:=D(L_2)\cap I^+(L_2)\cap I^-(T_+)$ and
$Q'_2:=\psi_2(Q_2)$, so that $Q_1\subset P_1^{\perp}$ and
$Q'_1\subset D(Q'_2)$.

So far the geometry of the proof. Now note that
$\mathcal{A}_{P_2}\subset\mathcal{A}_R$ by lemma
\ref{Cauchydevelopment} on $D(N_2)\cap I^+(N_2)$ and that
$\mathcal{A}_{R'}=\alpha_{\Psi_2}(\mathcal{A}_R)$. Applying
lemma \ref{Cauchydevelopment} in $D(N'_1)\cap I^-(N'_1)$ we see that
$\mathcal{A}_{R'}\subset\mathcal{A}_{P'_1}$ and we have
$\mathcal{A}_{P_1}=\alpha_{\Psi_1}^{-1}(\mathcal{A}_{P'_1})$.
Putting this together yields the inclusion:
\[
\alpha(\mathcal{A}_{P_2})\subset\alpha(\mathcal{A}_R)=
\alpha_{\Psi_1}^{-1}(\mathcal{A}_{R'})\subset
\alpha_{\Psi_1}^{-1}(\mathcal{A}_{P'_1})=\mathcal{A}_{P_1}.
\]
Similarly we have
$\mathcal{A}_{Q_1}=\alpha_{\Psi_1}^{-1}(\mathcal{A}_{Q'_1})$,
$\mathcal{A}_{Q'_2}=\alpha_{\Psi_2}(\mathcal{A}_{Q_2})$ and
$\mathcal{A}_{Q'_1}\subset\mathcal{A}_{Q'_2}$ by lemma
\ref{Cauchydevelopment}. This yields the inclusion:
\[
\alpha(\mathcal{A}_{Q_2})=\alpha_{\Psi_1}^{-1}(\mathcal{A}_{Q'_2})
\supset\alpha_{\Psi_1}^{-1}(\mathcal{A}_{Q'_1})=\mathcal{A}_{Q_1}.
\]
\end{proof*}

\section{The Reeh-Schlieder Property in Curved Spacetime}\label{RSsection}

The spacetime deformation argument of the previous section will
have some consequences for the Reeh-Schlieder property that we
describe in the current section. Unfortunately it is not clear that
we can deform a Reeh-Schlieder state into another (full) Reeh-Schlieder
state, but we do have the following more limited result:
\begin{theorem}\label{RSdeformation}
Consider a locally covariant quantum field theory $\mathbf{A}$ with
state space $\mathbf{S}$ which satisfies the time-slice axiom. Let $M_i$
be two globally hyperbolic spacetimes with diffeomorphic Cauchy surfaces
and suppose that $\omega_1\in S_{M_1}$ is a Reeh-Schlieder state. Then
given any bounded cc-region $O_2\subset\mathcal{M}_2$ with non-empty
causal complement, $O_2^{\perp}\not=\emptyset$, there is a
$^*$-isomorphism $\map{\alpha}{\mathcal{A}_{M_2}}{\mathcal{A}_{M_1}}$
such that $\omega_2:=\alpha^*(\omega_1)$ has the Reeh-Schlieder property
for $O_2$.

Moreover, if the Cauchy surfaces of the $M_i$ are non-compact and
$P_2\subset\mathcal{M}_2$ is a bounded cc-region, then there is a
bounded cc-region $Q_2\subset P_2^{\perp}$ for which $\omega_2$ has the
Reeh-Schlieder property.
\end{theorem}
\begin{proof*}
For the first statement let $\alpha$ and $U_1$ be as in the first part
of proposition \ref{isomorphism2} and note that $\alpha$ gives rise to a
unitary map
$\map{\mathtt{U}_{\alpha}}{\mathcal{H}_{\omega_2}}{\mathcal{H}_{\omega_1}}$.
This map is the expression of the essential uniqueness of the
GNS-representation, so that
$\mathtt{U}_{\alpha}\Omega_{\omega_2}=\Omega_{\omega_1}$ and
$\mathtt{U}_{\alpha}\pi_{\omega_2}\mathtt{U}_{\alpha}^*=\pi_{\omega_1}
\circ\alpha$.
The Reeh-Schlieder property for $O_2$ then follows from the observation
that $\mathtt{U}_{\alpha}\pi_{\omega_2}(\mathcal{A}_{O_2})
\mathtt{U}_{\alpha}^*\supset\pi_{\omega_1}(\mathcal{A}_{U_1})$:
\[
\overline{\pi_{\omega_2}(\mathcal{A}_{O_2})\Omega_{\omega_2}}\supset
\overline{\mathtt{U}_{\alpha}^*\pi_{\omega_1}(\mathcal{A}_{U_1})\Omega_{\omega_1}}
=\mathtt{U}_{\alpha}^*\mathcal{H}_{\omega_1}=
\mathcal{H}_{\omega_2}.
\]

Similarly for the second statement, given a bounded cc-region $P_2$ and
choosing $Q_1,Q_2$ as in the second statement of proposition
\ref{isomorphism2} we see that
$\mathtt{U}_{\alpha}\pi_{\omega_2}(\mathcal{A}_{Q_2})
\mathtt{U}_{\alpha}^*\supset\pi_{\omega_1}(\mathcal{A}_{Q_1})$.
\end{proof*}
The second part of theorem \ref{RSdeformation} means that $\omega_2$
is a Reeh-Schlieder state for all cc-regions that are big enough.
Indeed, if $V_2$ is a sufficiently small cc-region then $V_2^{\perp}$ is
connected (recall that we work with four-dimensional spacetimes) and
therefore $\omega_2$ has the Reeh-Schlieder property for some
cc-region in $V_2^{\perp}$ and hence also for $V_2^{\perp}$ itself.

A useful consequence of theorem \ref{RSdeformation} is the following:
\begin{corollary}\label{separability}
In the situation of theorem \ref{RSdeformation} if $\mathbf{A}$ is
causal then $\Omega_{\omega_2}$ is a cyclic and separating vector for
$\mathcal{R}^{\omega_2}_{O_2}$. If the Cauchy surfaces are non-compact
$\Omega_{\omega_2}$ is a separating vector for all
$\mathcal{R}^{\omega_2}_{P_2}$ where $P_2$ is a bounded cc-region.
\end{corollary}
\begin{proof*}
Recall that a vector is a separating vector for a von Neumann algebra
$\mathcal{R}$ iff it is a cyclic vector for the commutant $\mathcal{R}'$
(see \cite{Kadison+} proposition 5.5.11.). Choosing $V_1$ as in the
first part of proposition \ref{isomorphism2} we have
$\mathtt{U}_{\alpha}\pi_{\omega_2}(\mathcal{A}_{O_2})
\mathtt{U}_{\alpha}^*\subset\pi_{\omega_1}(\mathcal{A}_{V_1})$ by the
inclusion (\ref{estimate1}). Therefore the commutant of
$\mathtt{U}_{\alpha}\mathcal{R}^{\omega_2}_{O_2}\mathtt{U}_{\alpha}^*$
contains $(\mathcal{R}^{\omega_1}_{V_1})'$. As
$V_1^{\perp}\not=\emptyset$ this commutant contains the local algebra of
some cc-region for which $\Omega_{\omega_1}$ is cyclic. Hence
$\Omega_{\omega_1}$ is a separating vector for
$\mathcal{R}^{\omega_1}_{V_1}$ and $\Omega_{\omega_2}$ for
$\mathcal{R}^{\omega_2}_{O_2}$.

If the Cauchy surfaces are non-compact, $P_2$ is a bounded region
and $Q_2$ is as in theorem \ref{RSdeformation}, then
$(\mathcal{R}^{\omega_2}_{P_2})'$ contains
$\pi_{\omega_2}(\mathcal{A}_{Q_2})$, for which $\Omega_{\omega_2}$ is
cyclic. It follows that $\Omega_2$ is separating for
$\mathcal{R}^{\omega_2}_{P_2}$.
\end{proof*}

If the state space is locally quasi-equivalent and large enough it is
possible to show the existence of full Reeh-Schlieder states. The proof
uses abstract existence arguments, as opposed to the proof of theorem
\ref{RSdeformation} which is constructive, at least in principle.
\begin{theorem}\label{RSdense}
Consider a locally covariant quantum field theory $\mathbf{A}$ with a
locally quasi-equivalent state space $\mathbf{S}$ which is causal and
satisfies the time-slice axiom. Assume that $\mathbf{S}$ is maximal in
the sense that for any state $\omega$ on some $\mathcal{A}_{M}$ which is
locally quasi-equivalent to a state in $S_M$ we have $\omega\in S_M$.

Let $M_i$, $i=1,2$, be two globally hyperbolic spacetimes with
diffeomorphic non-compact Cauchy surfaces and assume that $\omega_1$ is
a Reeh-Schlieder state on $M_1$. Then $S_{M_2}$ contains a (full)
Reeh-Schlieder state.
\end{theorem}
\begin{proof*}
Let $\left\{O_n\right\}_{n\in\mathbb{N}}$ be a countable cover of
$\mathcal{M}_2$ consisting of bounded cc-regions with non-empty causal
complement. We then apply theorem \ref{RSdeformation} to each $O_n$ to
obtain a sequence of states $\omega^{n}_2\in S_{M_2}$ which have the
Reeh-Schlieder property for $O_n$. We write $\omega:=\omega^{1}_2$
and let $(\pi,\Omega,\mathcal{H})$ denote its GNS-representation.

For all $n\ge 2$ we now find a bounded cc-region
$V_n\subset\mathcal{M}_2$ such that $V_n\supset O_1\cup O_n$. For this
purpose we first choose a Cauchy surface $C\subset\mathcal{M}_2$ and
note that $K_n:=C\cap J(\overline{O_n})$ is compact. Letting
$L_n\subset C$ be a compact connected set containing $K_1\cup K_n$ in
its interior it suffices to choose
$V_n:=\mathrm{int}(D(L_n))\cap I^-(C_+)\cap I^+(C_-)$ for Cauchy
surfaces $C_{\pm}$ to the future resp. past of $O_1$, $O_n$ and $C$.
Note that $\Omega$ and $\Omega_{\omega^{n}_2}$ are cyclic and separating
vectors for $\mathcal{R}^{\omega}_{V_n}$ and
$\mathcal{R}^{\omega^{n}_2}_{V_n}$ respectively by
$O_1\cup O_n\subset V_n$ and by corollary \ref{separability}.
Because $\omega$ and $\omega^n_2$ are locally quasi-equivalent there
is a $^*$-isomorphism $\map{\phi}{\mathcal{R}^{\omega^n_2}_{V_n}}
{\mathcal{R}^{\omega}_{V_n}}$. In the presence of the cyclic and
separating vectors $\phi$ is implemented by a unitary map
$\map{\mathtt{U}_n}{\mathcal{H}_{\omega^n_2}}{\mathcal{H}}$ (see
\cite{Kadison+} theorem 7.2.9). We claim that
$\psi_n:=\mathtt{U}_n\Omega_{\omega^n_2}$ is cyclic for
$\mathcal{R}^{\omega}_{O_n}$. Indeed, by the definition of
quasi-equivalence we have $\phi\circ\pi_{\omega^n_2}=\pi_{\omega}$ on
$\mathcal{A}_{V_n}$, so
\[
\overline{\pi_{\omega}(\mathcal{A}_{O_n})\psi_n}=
\overline{\mathtt{U}_n\pi_{\omega^n_2}(\mathcal{A}_{O_n})
\Omega_{\omega^n_2}}=\mathtt{U}_n\mathcal{H}_{\omega^n_2}=
\mathcal{H}_{\omega}.
\]

We now apply the results of \cite{Dixmier+} to conclude that
$\mathcal{H}$ contains a dense set of vectors $\psi$ which are cyclic and
separating for all $\mathcal{R}^{\omega}_{O_n}$ simultaneously. Because
each cc--region $O\subset\mathcal{M}_2$ contains some $O_n$ we see that
$\omega_{\psi}:A\mapsto\frac{\langle\psi,\pi_{\omega}(A)\psi\rangle}
{\|\psi\|^2}$ defines a full Reeh-Schlieder state. Finally, because the
GNS-representation of $\omega_{\psi}$ is just $(\pi,\psi,\mathcal{H})$
we see that it is locally quasi-equivalent to $\omega$ and hence
$\omega_{\psi}\in S_{M_2}$.
\end{proof*}

In situations of physical interest it remains to be seen whether the
state space is big enough to contain such Reeh-Schlieder states.
However, theorem \ref{RSdeformation} is already enough for some
applications, such as the following conclusion concerning the type of
local von Neumann algebras
\begin{corollary}\label{not_finite}
Consider a nowhere-classical causal locally covariant quantum field
theory $\mathbf{A}$ with a locally quasi-equivalent state space
$\mathbf{S}$ which satisfy the time-slice axiom. Let $M_i$ be two
globally hyperbolic spacetimes with diffeomorphic Cauchy surfaces and
let $\omega_1\in S_{M_1}$ be a Reeh-Schlieder state. Then for any state
$\omega\in S_{M_i}$ and any cc-region $O\subset\mathcal{M}_i$ the local
von Neumann algebra $\mathcal{R}^{\omega}_O$ is not finite.
\end{corollary}
\begin{proof*}
We will use proposition 5.5.3 in \cite{Baumgartel+}, which says that
$\mathcal{R}^{\omega}_O$ is not finite if the GNS-vector $\Omega$ is a
cyclic and separating vector for $\mathcal{R}^{\omega}_O$ and for a
proper sub-algebra $\mathcal{R}^{\omega}_V$. Note that we can drop
the superscript $\omega$ if $O$ and $V$ are bounded, by local
quasi-equivalence.

First we consider $M_1$. For any bounded cc-region
$O_1\subset\mathcal{M}_1$ such that $O_1^{\perp}\not=\emptyset$ we
can find bounded cc-regions $O'\subset O_1^{\perp}$ and
$U,V\subset O_1$ such that $U\subset V^{\perp}$. By the Reeh-Schlieder
property the GNS-vector $\Omega_{\omega_1}$ is cyclic for
$\mathcal{R}_V$ and hence also for $\mathcal{R}_{O_1}$. Moreover it is
cyclic for $\mathcal{R}_{O_1}'\supset\mathcal{R}_{O'}$ and therefore it is
separating for $\mathcal{R}_{O_1}$ and $\mathcal{R}_V$. Now suppose that
$\mathcal{R}_{O_1}=\mathcal{R}_V$. Then, by causality:
\[
\pi_{\omega}(\mathcal{A}_U)\subset \pi_{\omega}(\mathcal{A}_V)'=
\pi_{\omega}(\mathcal{A}_{O_1})'\subset 
\pi_{\omega}(\mathcal{A}_U)'.
\]
It follows that $\mathcal{R}_U\subset\mathcal{R}_U'$, which
contradicts the nowhere classicality. Therefore, the inclusion
$\mathcal{R}_V\subset\mathcal{R}_{O_1}$ must be proper and the cited
theorem applies. Of course, if
$O\subset\mathcal{M}_1$ is a cc-region that is not bounded, then it
contains a bounded sub-cc-region $O_1$ as above and
$\mathcal{R}^{\omega}_O\supset\mathcal{R}^{\omega}_{O_1}
\simeq\mathcal{R}_{O_1}$ isn't finite either for any $\omega\in
S_{M_1}$. (If $V$ is a partial isometry in the smaller algebra such that
$I=V^*V$ and $E:=VV^*<I$ then the same $V$ shows that $I$ is not finite
in the larger algebra.)

Next we consider $M_2$ and let $O\subset\mathcal{M}_2$ be any
cc-region. It contains a cc-region $O_2$ with
$O_2^{\perp}\not=\emptyset$,
so we can apply theorem \ref{RSdeformation}. Using the unitary map
$\map{\mathtt{U}_{\alpha}}{\mathcal{H}_{\omega_2}}{\mathcal{H}_{\omega_1}}$
we see that $\mathcal{R}_{O_2}\simeq\mathcal{R}^{\omega_2}_{O_2}$
contains $\alpha^{-1}(\mathcal{R}^{\omega_1}_{O_1})$, which is not
finite by the first paragraph. Hence $\mathcal{R}_{O_2}$ is not finite
and the statement for $O$ then follows again by inclusion.
\end{proof*}
Instead of the nowhere classicality we could have assumed that the local
von Neumann algebras in $M_1$ are infinite, which allows us to derive
the same conclusion for $M_2$. Unfortunately it is in general impossible
to completely derive the type
of the local algebras using this kind of argument. Even if we know the
types of the algebras $\mathcal{A}_{U_1}$ and $\mathcal{A}_{V_1}$ in the
inclusions (\ref{estimate1}), we can't deduce the type of
$\mathcal{A}_{O_2}$.

Another important consequence of proposition \ref{RSdeformation} is that
corollary \ref{separability} enables us to apply the Tomita-Takesaki
modular theory to $\mathcal{R}^{\omega_2}_{O_2}$ (or to the von Neumann
algebra of any bounded cc-region $V_2$ which contains $O_2$, if the
Cauchy surfaces are non-compact). More precisely, let
$O_2\subset\mathcal{M}_2$ be given and let $U_1,V_1\subset\mathcal{M}_1$
be the bounded cc-regions and
$\map{\alpha}{\mathcal{M}_2}{\mathcal{M}_1}$ the $^*$-isomorphism of
proposition \ref{isomorphism2}, so that $\mathcal{A}_{O_1}\subset
\alpha(\mathcal{A}_{O_2})\subset\mathcal{A}_{V_1}$. We can then define
$\mathcal{R}:=\mathtt{U}_{\alpha}\mathcal{R}^{\omega_2}_{O_2}
\mathtt{U}_{\alpha}^*$ and obtain $\mathcal{R}^{\omega_1}_{U_1}\subset
\mathcal{R}\subset\mathcal{R}^{\omega_1}_{V_1}$. It is then clear that
the respective Tomita-operators are extensions of each other,
$S_{U_1}\subset S_{\mathcal{R}}\subset S_{V_1}$ (see e.g.
\cite{Kadison+}).

\section{The free scalar field}\label{freefieldsection}

As an example we will consider the free scalar field, which can be
quantised using the Weyl algebra (see \cite{Dimock}). For a globally
hyperbolic spacetime $M$ the algebra $\mathcal{A}_M$ is defined as
follows. We let $E:=E^+-E^-$
denote the difference of the advanced and retarded fundamental solution
of the Klein-Gordon operator $\nabla^a\nabla_a+m^2$ for a given mass
$m\ge 0$. The linear space $H:=E(C_0^{\infty}(\mathcal{M}))$ has a
non-degenerate symplectic form defined by
$\sigma(Ef,Eg):=\int_{\mathcal{M}}fEg$, where we integrate with respect
to the volume element determined by the metric. To every $Ef\in H$ we
can then associate an element $W(Ef)$ subject to the relations
\[
W(Ef)^*=W(-Ef),\quad W(Ef)W(Eg)=e^{-\frac{i}{2}\sigma(Ef,Eg)}W(E(f+g)).
\]
These elements form a $^*$-algebra that can be given a norm and
completed to a $C^*$-algebra $\mathcal{A}_M$. It is shown in
\cite{Brunetti+} theorem 2.2 that the scalar free field is an example of
a locally covariant quantum field theory which is causal. It satisfies
part of the time-slice axiom, namely if $O\subset M$ contains a Cauchy
surface then $\mathcal{A}_O=\mathcal{A}_M$.\footnote{Note that this is
what \cite{Brunetti+} call the time slice axiom. In our definition,
however, we also need to choose a suitable state space functor so that
we get isomorphisms of the sets of states too.}

A state $\omega$ on $\mathcal{A}_M$ is called regular if the group of
unitary operators $\lambda\mapsto\pi_{\omega}(W(\lambda Ef))$ is
strongly continuous for each $f$. It then has a self-adjoint (unbounded)
generator $\Phi_{\omega}(f)$ and we can define the Hilbert-space valued
distribution $\phi_{\omega}(f):=\Phi_{\omega}(f)\Omega_{\omega}$. A
regular state is quasi-free iff the two-point function
\[
w_2(f,h):=\langle\phi_{\omega}(\bar{f}),\phi_{\omega}(h)\rangle,
\quad f,h\in C_0^{\infty}(\mathcal{M})
\]
determines the state by $\omega(W(Ef))=e^{-w_2(f,f)}$. A quasi-free
state is Hadamard iff
$WF_{\infty}(\phi_{\omega}(.))\subset\overline{V^+}$, where
$\overline{V^+}\subset T^*\mathcal{M}$ denotes the cone of future
directed causal covectors of the spacetime (see \cite{Strohmaier+}
proposition 6.1). Quasi-free Hadamard states exist on all globally
hyperbolic spacetimes (see \cite{Fulling+}) and they are believed to be
the most suitable states to play a role similar to the vacuum in
Minkowski spacetime. For this reason we will want to choose a state
space $S_M$ which contains
all quasi-free Hadamard states. If we choose these states only it can
be shown that we get a locally quasi-equivalent state space (see
\cite{Verch2} theorem 3.6) and the time-slice axiom is satisfied
(see \cite{Radzikowski} theorem 5.1 and the subsequent discussion).

We may now apply the results of section \ref{RSsection}:
\begin{proposition}\label{freefield}
Let $M$ be a globally hyperbolic spacetime, let $O\subset\mathcal{M}$ a
bounded cc-region with non-empty causal complement and assume
that the mass $m>0$ is strictly positive. Then there is a Hadamard state
$\omega$ on $\mathcal{A}_M$ which has the Reeh-Schlieder property for
$O$. The vector $\Omega_{\omega}$ is cyclic and separating for
$\mathcal{R}_O$. For all bounded cc-regions $V\subset\mathcal{M}$ the
local von Neumann algebra $\mathcal{R}_V$ is not finite. Moreover, if
the Cauchy surfaces of $M$ are non-compact then $\Omega_{\omega}$ is a
separating vector for all $\mathcal{R}_V$.
\end{proposition}
\begin{proof*}
The theory is causal, satisfies the time-slice axiom and the state space
is locally quasi-equivalent. Moreover, the theory is nowhere classical.
To see this we note that the local $C^*$-algebras are non-commutative
and simple, so the representations $\pi_{\omega}$ are faithful. Now we
can find an ultrastatic (and hence stationary) spacetime $M'$
diffeomorphic to $M$. Because $m>0$ we may apply the results of
\cite{Kay}, which imply the existence of a regular quasi-free ground
state $\omega'$ on $M'$. This state has the Reeh-Schlieder property (see
\cite{Strohmaier}) and is Hadamard because it satisfies the microlocal
spectrum condition (see \cite{Strohmaier+,Radzikowski}). The conclusions
now follow immediately from theorem \ref{RSdeformation} and the
corollaries \ref{separability} and \ref{not_finite}. Note that stronger
results on the type of the local algebras are known, \cite{Verch2}.
\end{proof*}

If we would enlarge our state space and allow any state that is locally
quasi-equivalent to a quasi-free Hadamard state, then it follows from
theorem \ref{RSdense} that it also contains full Reeh-Schlieder states.
In fact, if $\omega$ is a suitable quasi-free Hadamard state on
$\mathcal{A}_M$ then the proof of theorem \ref{RSdense} shows that
$\mathcal{H}_{\omega}$ contains a dense $G_{\delta}$ of vectors which
define Reeh-Schlieder states. An important question is how many states
are both Hadamard and Reeh-Schlieder states. As a partial answer we wish
to note the following. If a vector $\psi\in\mathcal{H}_{\omega}$ defines
a Hadamard state then it must be in the domain of the unbounded
self-adjoint operator $\Phi_{\omega}(f)$ for some real-valued test
function $f$. We then apply
\begin{proposition}\label{domain}
The domain of an unbounded self-adjoint operator $T$ on a Hilbert space
$\mathcal{H}$ is a meagre $F_{\sigma}$, (i.e. the complement of a dense
$G_{\delta}$).
\end{proposition}
\begin{proof*}
For each $n\in\mathbb{N}$ we define
$V_n:=\left\{\psi\in\mathcal{H}| \|T\psi\|\le n\right\}$ and note that
$\mathrm{dom}(T)=\cup_n V_n$. The sets $V_n$ are nowhere dense because
$T$ is unbounded. They are also closed because for a Cauchy sequence
$\psi_i\rightarrow\psi$ with $\psi_i\in V_n$ we have
$\|TE_{[-r,r]}\psi\|\le\|TE_{[-r,r]}(\psi-\psi_i)\|+
\|TE_{[-r,r]}\psi_i\|\le r\|\psi-\psi_i\|+n$, where $E_{[-r,r]}$ is the
spectral projection of $T$ on the interval $[-r,r]$. Taking
$i\rightarrow\infty$ shows that $\|TE_{[-r,r]}\psi\|\le n$ for all $r$
and hence $\|T\psi\|\le n$, i.e. $\psi\in V_n$. This completes the
proof.
\end{proof*}
It follows that most Reeh-Schlieder states are not Hadamard. The
converse question, how many Hadamard states are Reeh-Schlieder states,
remains open.

\section{Conclusions}\label{conclusions}

If one accepts locally covariant quantum field theory as a suitable
axiomatic framework to describe quantum field theories in curved
spacetime then one only needs to assume the very natural time-slice
axiom in order to use the general technique of spacetime deformation.
The geometrical ideas behind deformation results like proposition
\ref{deformation2} are insightful, even though the proofs can become
a bit involved. It should be noted, however, that these geometrical
results, possibly combined with other assumptions such as
causality, have immediate consequences on the algebraic side which
are not hard to prove. This we have seen in section \ref{RSsection},
where most proofs follow easily from the deformation, with the
exception of theorem \ref{RSdense}.

Concerning the Reeh-Schlieder property we have shown that a
Reeh-Schlieder state on one spacetime can be deformed in such a way
that it gives a state on a diffeomorphic spacetime which is a
Reeh-Schlieder state for a given cc-region. It is even possible to get
full Reeh-Schlieder states, but it is not clear whether these are
``physical'' enough to belong to a state space of interest.
Nevertheless, our results ado llow us to draw conclusions about the
type of local von Neumann algebras and they open up the way to use
Tomita-Takesaki theory in curved spacetime.

${}$\\[15pt]
{\bf Acknowledgements}\\[6pt]
I would like to thank Chris Fewster for suggesting the current approach
to the Reeh-Schlieder property and for many helpful discussions and
comments on the second draft. Many thanks also to Lutz Osterbrink for
his careful proofreading of the first draft.


\thebibliography{              }

\bibitem{Baumgartel+}
Baumg\"artel, H. and Wollenberg, M.,
\emph{Causal nets of operator algebras},
Akademie Verlag, Berlin (1992)

\bibitem{Bernal+1}
Bernal, A.N. and S\'anchez, M.,
\emph{Smoothness of time functions and the metric splitting of globally
hyperbolic spacetimes},
Commun. Math. Phys. \textbf{257}, 43--50 (2005)

\bibitem{Bernal+2}
Bernal, A.N. and S\'anchez, M.,
\emph{Further results on the smoothability of Cauchy hypersurfaces and
Cauchy time functions},
Lett. Math. Phys. \textbf{77}, 183--197 (2006)

\bibitem{Brunetti+}
Brunetti, R., Fredenhagen, K. and Verch, R.,
\emph{The generally covariant locality principle---a new paradigm for
local quantum field theory},
Commun. Math. Phys. \textbf{237}, 31--68 (2003)

\bibitem{Brunetti+2}
Brunetti, R. and Ruzzi, G.,
\emph{Superselection sectors and general covariance. I},
Commun. Math. Phys. \textbf{270}, 69--108 (2007)

\bibitem{Dimock}
Dimock, J.,
\emph{Algebras of local observables on a manifold},
Commun. Math. Phys. \textbf{77}, 219--228 (1980)

\bibitem{Dixmier+}
Dixmier, J. and Mar\'echal, O.,
\emph{Vecteurs totalisateurs d'une alg\`ebre de von Neumann},
Commun. Math. Phys. \textbf{22}, 44--50 (1971)

\bibitem{Fulling+}
Fulling, S.A., Narcowich, F.J. and Wald, R.M.,
\emph{Singularity structure of the two-point function in quantum field
theory in curved spacetime, II},
Ann. Phys. (N.Y.) \textbf{136}, 243--272 (1981)

\bibitem{Haag}
Haag, R.,
\emph{Local quantum physics -- fields, particles, algebras},
Springer Verlag Berlin-Heidelberg, (1992)

\bibitem{Hawking+}
Hawking, S.W. and Ellis, G.F.R.,
\emph{The large scale structure of space-time},
Cambridge University Press, Cambridge, (1973)

\bibitem{Kadison+}
Kadison, R.V. and Ringrose, J.R.,
\emph{Fundamentals of the theory of operator algebras}
Academic Press, London, (1983)

\bibitem{Kay}
Kay, B.S.,
\emph{Linear spin-zero quantum fields in external
gravitational and scalar fields. I. A one particle structure for the
stationary case},
Commun. Math. Phys. \textbf{62}, 55--70 (1978)

\bibitem{ONeill}
O'Neill, B.,
\emph{Semi-Riemannian geometry: with applications to relativity},
Academic Press, New York (1983)

\bibitem{Radzikowski}
Radzikowski, M.J.,
\emph{Micro-local approach to the Hadamard condition in quantum
field theory on curved space-time},
Commun. Math. Phys. \textbf{179}, 529--553 (1996)

\bibitem{Reeh+}
Reeh, H. and Schlieder, S.,
\emph{Bemerkungen zur Unit\"ar\"aquivalenz von Lorentzinvarianten
Felden},
Nuovo Cimento \textbf{22}, 1051--1068 (1961)

\bibitem{Strohmaier}
Strohmaier, A.,
\emph{The Reeh-Schlieder property for quantum fields on stationary
spacetimes},
Commun. Math. Phys. \textbf{215}, 105--118 (2000)

\bibitem{Strohmaier+}
Strohmaier, A., Verch, R. and Wollenberg, M.,
\emph{Microlocal analysis of quantum fields on curved space-times: analytic
wavefront sets and Reeh-Schlieder theorems},
J. Math. Phys. \textbf{43}, 5514--5530 (2002)

\bibitem{Verch1}
Verch, R.,
\emph{A spin-statistics theorem for quantum fields on curved spacetime
manifolds in a generally covariant framework},
Commun. Math. Phys. \textbf{223}, 261--288 (2001)

\bibitem{Verch2}
Verch, R.,
\emph{Continuity of symplectically adjoint maps and the algebraic
structure of Hadamard vacuum representations for quantum fields on
curved spacetime},
Rev. Math. Phys. \textbf{9}, 635--674 (1997)

\bibitem{Wald}
Wald, R.M.,
\emph{General relativity},
The University of Chicago Press, Chicago and London, (1984)
\end{document}